\documentclass[aps,prl,twocolumn,preprintnumbers,10pt,showpacs,nofootinbib]{revtex4-1}

\usepackage{amsmath,bm,amssymb}

\usepackage{graphicx}
\usepackage{psfrag}
\usepackage[caption=false]{subfig}

\usepackage{color,xcolor}
\definecolor{urlblue}{rgb}{0.2,0.4,0.7}
\definecolor{citegreen}{rgb}{0,0.6,0.2}
\definecolor{linkred}{rgb}{0.9,0.2,0.1}
\usepackage{hyperref}
\hypersetup{
colorlinks=true, citecolor=citegreen, linkcolor=blue,urlcolor=urlblue}

\usepackage{slashed}

\usepackage{tikz}
\usetikzlibrary{positioning,arrows}
\usetikzlibrary{decorations.pathmorphing}
\usetikzlibrary{decorations.markings}
\usetikzlibrary{shapes.geometric}
\tikzset{
    vector/.style={decorate, decoration={snake}, draw},
    provector/.style={decorate, decoration={snake,amplitude=2.5pt}, draw},
    antivector/.style={decorate, decoration={snake,amplitude=-2.5pt}, draw},
    fermion/.style={draw=black,
      postaction={decorate},decoration={markings,mark=at position .55
        with {\arrow[draw=black]{>}}}}, 
    fermionbar/.style={draw=black, postaction={decorate},
                       decoration={markings,mark=at position .55 with {\arrow[draw=black]{<}}}},
    fermionnoarrow/.style={draw=black},
    gluon/.style={decorate, draw=black,decoration={coil,amplitude=4pt, segment length=4pt}},
    scalar/.style={dashed,draw=black,
      postaction={decorate},decoration={markings,mark=at position .55
        with {\arrow[draw=black]{>}}}}, 
    scalarbar/.style={dashed,draw=black,
      postaction={decorate},decoration={markings,mark=at position .55
        with {\arrow[draw=black]{<}}}}, 
    scalarnoarrow/.style={dashed,draw=black},
    electron/.style={draw=black,
      postaction={decorate},decoration={markings,mark=at position .55
        with {\arrow[draw=black]{>}}}}, 
    bigvector/.style={decorate, decoration={snake,amplitude=4pt}, draw},
}

\def\gm{\gamma}

\def\ep{\epsilon}
\def\epm1{\frac{1}{\epsilon}}
\def\epm2{\frac{1}{\epsilon^{2}}}
\def\epm3{\frac{1}{\epsilon^{3}}}
\def\epm4{\frac{1}{\epsilon^{4}}}

\def\bt{\beta}

\def\bt0{\beta_{0}}
\def\bt1{\beta_{1}}
\def\bt2{\beta_{2}}
\def\bt3{\beta_{3}}
\def\gm0{\gamma_{0}}
\def\gm1{\gamma_{1}}
\def\gm2{\gamma_{2}}
\def\gm3{\gamma_{3}}

\def\l{\left}
\def\r{\right}

\def\ep{\epsilon}

\def\qgraf{{\fontfamily{qcr}\selectfont
QGRAF}}
\def\python{{\fontfamily{qcr}\selectfont
PYTHON}}
\def\form{{\fontfamily{qcr}\selectfont
FORM}}

\def\litered{{\fontfamily{qcr}\selectfont
LiteRed}}

\def\arXiv{{\fontfamily{qcr}\selectfont
arXiv}}

\def\polylogtools{{\fontfamily{qcr}\selectfont
PolyLogTools}}

\def\ancillary{{\fontfamily{qcr}\selectfont
ancillary}}

\allowdisplaybreaks

\begin{document}

\def\l{\left}
\def\r{\right}
\def\ep{\epsilon}
\def\bt{\beta}

\preprint{IMSC/2019/11/11, MPP-2019-232, PSI-PR-19-23}

\title{Form factors with two operator insertions and the principle of maximal transcendentality}

\author{Taushif Ahmed$^{a}$}\email{taushif@mpp.mpg.de}
\author{Pulak Banerjee$^{b}$}\email{pulak.banerjee@psi.ch}
\author{Amlan Chakraborty$^{c}$}\email{amlanchak@imsc.res.in}
\author{Prasanna K. Dhani$^{d}$}\email{prasannakumar.dhani@fi.infn.it}
\author{V. Ravindran$^{c}$}\email{ravindra@imsc.res.in}

\affiliation{\rm $^a$Max-Planck-Institut f\"ur Physik, Werner-Heisenberg-Institut, 80805 M\"unchen, Germany\\
$^b$Paul Scherrer Institut, CH-5232 Villigen PSI, Switzerland \\ 
$^c$The Institute of Mathematical Sciences, HBNI, Taramani, Chennai 600113, India \\ 
$^d$INFN, Sezione di Firenze, I-50019 Sesto Fiorentino, Florence, Italy} 

\begin{abstract}
\noindent
We present the first calculations of two-point two-loop form factors (FFs) with a two identical operators insertion in maximally supersymmetric Yang-Mills theory. In this article, we consider the supersymmetry protected half-BPS primary and unprotected Konishi operators. Unlike the FFs of a single operator insertion of the half-BPS primary, the FFs involving two half-BPS operators are found to contain lower transcendentality weight terms in addition to the highest ones. Moreover, in contrast to Sudakov FFs, the highest weight terms of the FFs of a double half-BPS no longer match with that of a double Konishi. We also find that the principle of maximal transcendentality, which dictates the presence of identical highest weight terms in the scalar FFs of half-BPS and quark/gluon FFs in QCD, does not hold true anymore for insertions of two identical operators. We discover the absence of any additional ultraviolet counterterm that could arise from the contact interaction between two composite operators.

\end{abstract}

\maketitle
\section{I.\,\,INTRODUCTION}

A generic quantum field theory is entirely specified by the knowledge of on shell scattering amplitudes and off shell correlation functions. There exists another class of fascinating objects, called form factors (FFs), which interpolate between amplitudes and correlators. This object is defined through the overlap between a state created by the action of composite gauge invariant operators on the vacuum and a state consisting of only on shell particles. The FFs in the maximally supersymmetric Yang-Mills theory (${\cal N}=4$ sYM) are expected to inherit much of the remarkable simplicity of the on shell amplitudes, and at the same time, to reflect some of the nontrivial behavior of the off shell correlators. In past few decades, FFs have been studied extensively starting from the seminal works in Refs.~\cite{vanNeerven:1985ja,Alday:2007he,Brandhuber:2010ad,Bork:2010wf,Brandhuber:2011tv,Bork:2011cj}. Very recently, the first step is taken to go beyond the horizon of FFs with a single operator insertion and the scenario with the two operator insertion is addressed in Ref.~\cite{Ahmed:2019upm}. In this article, we take this step forward by performing a state-of-the-art computation to explore the nature of two-loop two-point FFs with insertions of two identical operators. Consequently, for the first time, we examine the validation of several conjectures in view of generalized FFs.

In this work, we consider two local gauge invariant operators,
\begin{align}
\label{eq:op-def}
&{\cal O}^{\rm{BPS}}_{rt} = \phi^b_r\phi^b_t - \frac{1}{3}\delta_{rt}\phi^b_s\phi^b_s\,,
\quad{\cal O}^{\cal K} &&= \phi^b_r\phi^b_r + \chi^b_r\chi^b_r,
\end{align}
%
where ${\cal O}^{\rm{BPS}}_{rt}$ and ${\cal O}^{\cal K}$ are the SUSY protected half-Bogomonlyi-Prasad-Sommerfeld (BPS) primary belonging to the stress-energy supermultiplet and unprotected Konishi operators, respectively. The scalar and pseudoscalar fields are denoted by $ \phi^b_r$ and $\chi^b_s$, respectively, where their number of generations is represented through $r, s, t \in [1,n_g]$ with $n_g=3$ in four dimensions. All the fields in the ${\cal N}=4$ sYM theory transform under adjoint representation, which is represented through the SU(N) color index $b$.

Understanding the analytical structures of on shell amplitudes and FFs in ${\cal N}=4$ sYM has been an active area of investigation, not only to uncover the hidden structures of these quantities but also to establish the connections with other gauge theories, such as QCD. One of the most intriguing facts is the appearance of uniform transcendental\footnote{The transcendentality weight, $\tau$, of a function, $f$, is defined as the number of iterated integrals required to define it, e.g., $\tau(\log)=1\,, \tau({\rm Li}_n)=n\,, \tau(\zeta_n)=n,$ and moreover, we define $\tau(f_1f_2)=\tau(f_1)+\tau(f_2)$. Algebraic factors are assigned to weight zero and the dimensional regularization parameter $\epsilon$ to -1.} (UT) weight terms in a certain class of quantities in ${\cal N}=4$ sYM. This is indeed an observational~\cite{Kotikov:2002ab,Kotikov:2004er,Bern:2006ew,Bork:2010wf,Gehrmann:2011xn,Brandhuber:2012vm,Eden:2012rr,Drummond:2013nda,Basso:2015eqa,Banerjee:2016kri,Banerjee:2018yrn}, albeit unproven fact. The two-point or Sudakov FFs of primary half-BPS operator belonging to the stress-energy supermultiplet is observed~\cite{vanNeerven:1985ja,Brandhuber:2010ad,Gehrmann:2011xn} to exhibit the UT property to three loops; more specifically, they are composed of only highest transcendental (HT) terms with weight 2L at loop order L. This is a consequence of the existence of an integral representation of the FFs with every Feynman integral as UT~\cite{Gehrmann:2011xn}. Knowing the existence of such a basis has profound implications in choosing the basis of integrals while evaluating Feynman integrals using a differential equations method~\cite{Henn:2013pwa}. The three-point FFs of half-BPS operator are also found to respect this wonderful UT property~\cite{Brandhuber:2012vm}. On the contrary, this property fails for the two-~\cite{Nandan:2014oga,Ahmed:2016vgl} and three-point~\cite{Banerjee:2016kri} FFs of the unprotected Konishi operator, which are investigated up to three and two loops, respectively. Three-point FFs of a Konishi descendant operator are also found not to exhibit the UT property~\cite{Ahmed:2019nkj}. All the aforementioned results are in accordance with the general belief that the FFs of a supersymmetry (SUSY) protected operator, such as the half-BPS primary,  exhibit UT behavior. Having seen the beautiful property of UT in FFs of an one operator insertion, the question arises whether it is respected for the two-point FFs with the SUSY protected two operator insertion and whether this property can be extrapolated to generalized FFs with a $n$-number of operators insertion. In this article, for the first time, we address this question, and we find that the UT property does not hold true at two loop for the FF of double half-BPS insertion.

It is conjectured in Ref.~\cite{Loebbert:2015ova} that the HT weight parts of every two-point minimal FFs (the presence of an equal number of fields in the operator and external on shell state) are identical, and those are equal to that of a half-BPS, ${\cal O}^{\rm BPS}_{rt}$. This conjecture is verified to four-loops order in Ref.~\cite{Ahmed:2016vgl} for the Sudakov FFs of operator ${\cal O}^{\cal K}$. Naturally, it is curious to see if this conjecture holds true for the generalized two-point FFs with the two operator insertion. In particular, we address whether the HT parts of the two-point FFs with a double ${\cal O}^{\rm BPS}_{rt}$ and double ${\cal O}^{\cal K}$ insertion match. It turns out they are different both at one and two loop. 


The connection between quantities in ${\cal N}=4$ sYM and that of QCD is of fundamental importance. In addition to deepening our theoretical understanding, it is motivated from the fact that computing a quantity in QCD is much more difficult, and in the absence of our ability to calculate a quantity in QCD, it is possible to obtain the result, at least partially, from that of simpler theory, such as ${\cal N}=4$ sYM. In Refs.~\cite{Kotikov:2001sc,Kotikov:2004er,Kotikov:2006ts,Banerjee:2018yrn}, it is found that the anomalous dimensions of the leading twist-two-operators in ${\cal N}=4$ sYM are identical to the HT counterparts in QCD~\cite{Moch:2004pa}, and consequently, the principle of maximal transcendentality (PMT) is conjectured. The PMT says that the algebraically most complex part of certain quantities in ${\cal N}=4$ sYM and QCD are identical. The conjecture is found to hold true for two-point FFs to three-loops level~\cite{Gehrmann:2011xn}; more specifically, the HT pieces of quark and gluon FFs in QCD~\cite{Gehrmann:2011aa} are identical, up to a normalization
factor of $2^L$, to scalar FFs of the operator ${\cal O}^{\rm BPS}_{rt}$ in ${\cal N}=4$ sYM upon changing the representation of fermions in QCD from fundamental to adjoint. The diagonal elements of the two-point pseudoscalar~\cite{Ahmed:2015qpa} and tensorial FFs~\cite{Ahmed:2015qia,Ahmed:2016qjf} also obey the conjecture. The three-point scalar and pseudoscalar FFs are also found to respect the PMT~\cite{Brandhuber:2012vm,Loebbert:2015ova,Loebbert:2016xkw,Banerjee:2017faz,Brandhuber:2017bkg,Jin:2018fak,Brandhuber:2018xzk,Jin:2019ile,Jin:2019opr}. 
Employing this conjecture, the four-loop collinear anomalous dimension in the planar ${\cal N}=4$ sYM is computed~\cite{Dixon:2017nat}. In Ref.~\cite{Belitsky:2013ofa}, the asymptotic limit of the energy-energy correlator, and in Ref.~\cite{Banerjee:2018yrn}, the soft function, are also observed to be consistent with PMT. However, the complete domain of validity of this principle is still unknown. For on shell amplitudes, it fails even at one loop~\cite{Bern:1993mq} with four or five external gluons. In this article, we investigate whether the wonderful conjecture of PMT holds true for two-point FFs of a two operator insertion with ${\cal O}^{\rm BPS}_{rt}$. Surprisingly, we find that the PMT also does not hold true.

\section{II.\,\,COMPUTATION OF TWO-LOOP FORM FACTOR}
\label{sec:framework}
The Lagrangian density~\cite{Brink:1976bc,Gliozzi:1976qd, Jones:1977zr,Poggio:1977ma} encapsulating the dynamics of ${\cal N}=4$ sYM and describing the interactions with the gauge invariant local operators, ${\cal O}^{\rm BPS}_{rt}$ and ${\cal O}^{\cal K}$, is given by
\begin{align}
\label{eq:lagrangian}
{\cal L} = {\cal L}_{\tiny{{\cal N}=4}} + {\cal J}^{\rm BPS}_{rt}{\cal O}^{\rm BPS}_{rt} + {\cal J}^{\cal K} {\cal O}^{\cal K}.
\end{align}
The quantity ${\cal J}$ represents the off shell state described by the corresponding operator. We are interested in investigating the two-point FFs with the two operator insertion of the following scattering processes:
\begin{align}
\label{eq:process}
    \phi(p_1)+\phi(p_2) &\rightarrow 
    \begin{cases}
    &{\cal J}^{\rm BPS}(p_3)+{\cal J}^{\rm BPS}(p_4)\,,\\
    &{\cal J}^{\cal K}(p_3)+{\cal J}^{\cal K}(p_4)\,,
    \end{cases}
\end{align}
where $p_i$ are the corresponding four momentum with $p_1^2=p_2^2= 0$ and $p_3^2=p_4^2=m_{\lambda}^2 \neq 0$. The $m^2_{\lambda}$ is the invariant mass square of the colour singlet state described by the operators in \eqref{eq:op-def}, i.e., $\lambda \in \{ {\rm BPS}, {\cal K}\}$. The underlying Mandelstam variables are defined as $s \equiv (p_1+p_2)^2$, $t \equiv (p_1-p_3)^2$ and $u \equiv (p_2-p_3)^2$ satisfying $s+t+u=2 m_{\lambda}^2$. For convenience, we introduce the dimensionless variables $x,y,z$ through
\begin{align}
\label{eq:xyz}
	s = m_{\lambda}^2 {(1+x)^2 \over x},\quad t = -m_{\lambda}^2 y,\quad u= -m_{\lambda}^2 z.
\end{align}
The quantity $x$ is called the Landau variable. The aforementioned choice of variables are necessary to rationalize the square root, $\sqrt{s(s-4m_{\lambda}^2)}$, which appears in the differential equation. The square root is associated to the threshold of two massive particles production. 

\begin{figure}[htb]
\begin{center}
\includegraphics[scale=0.55]{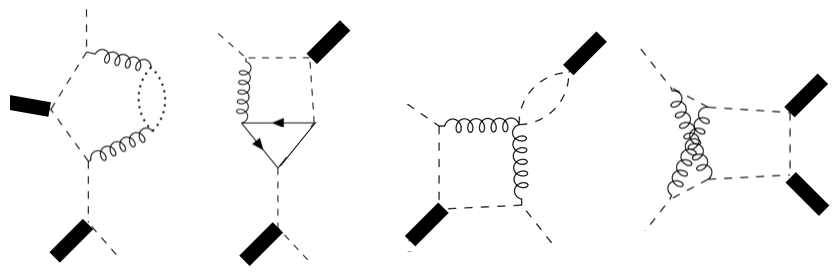}
\caption{Sample of two-loop Feynman diagrams. The thick solid, thin solid, dashed, dotted, and curly lines represent ${\cal J}^{\lambda}$, Majorana fermion, scalar, ghost, and gluon, respectively.}
\label{dia:2L}
\end{center}
\end{figure}

In perturbation theory, the scattering amplitude of the processes \eqref{eq:process} can be expanded in powers of the 't Hooft coupling constant, $a \equiv {g^2 N} (4 \pi e^{-\gamma_E})^{-\frac{\epsilon}{2}}/{(4\pi)^2}$, as $|{\cal M} \rangle_{\lambda} = \sum_{n=0}^{\infty} a^n |{\cal M}^{(n)}\rangle_{\lambda}$
%
%
where the quantity $|{\cal M}^{(n)}\rangle_{\lambda}$ represents the $n$th loop amplitude of the process involving ${\cal J}^{\lambda}$. The quadratic Casimir in the adjoint representation of SU(N) group is given by $N$. The dimensional regulator, $\epsilon$, is defined through $d=4+\epsilon$ with the space-time dimension $d$. We regulate the theory by adopting a SUSY preserving modified dimensional reduction (${\overline{\rm DR}}$) scheme~\cite{Siegel:1979wq,Capper:1979ns}, which keeps the number of bosonic and fermionic degrees of freedom equal. This is achieved by changing the number of scalar and pseudoscalar generations  from $n_g=3$ to $n_{g,\epsilon}=3-\epsilon/2$ in $d$ dimensions. The FFs are constructed out of the transition matrix elements through
\begin{align}
\label{eq:FF-defn}
    {\cal F}_{\lambda} = 1+\sum_{n=1}^{\infty} a^n {\cal F}^{(n)}_{\lambda} \equiv 1+\sum_{n=1}^{\infty} a^n  \frac{\langle {\cal M}^{(0)}|{\cal M}^{(n)} \rangle }{\langle {\cal M}^{(0)}|{\cal M}^{(0)} \rangle}\bigg\vert_{\lambda}\,.
\end{align}
The primary objective of this article is to compute the FFs to two loops, i.e., ${\cal F}^{(1)}_{\lambda}$ and ${\cal F}^{(2)}_{\lambda}$. In Fig.~\ref{dia:2L}, we show some examples of two-loop Feynman diagrams.

In contrast to the widely used method of unitarity to evaluate the on shell amplitudes and FFs in ${\cal N}=4$ sYM, we employ the methodology based on Feynman diagrammatic approach. The relevant Feynman diagrams are generated using \qgraf~\cite{Nogueira:1991ex}. Because of the presence of Majorana fermions in the theory, the generated diagrams are plagued with the wrong flow of fermionic currents which is rectified by an in-house algorithm based on \python. There are 440 and 606 number of Feynman diagrams at two loop for the production of double ${\cal J}^{\rm BPS}_{rt}$ and ${\cal J}^{\cal K}$, respectively. The diagrams are passed through a series of in-house codes based on symbolic manipulating program \form~\cite{Vermaseren:2000nd} in order to apply the Feynman rules, perform spinor, Lorentz, and color algebra. To ensure the inclusion of only physical degrees of freedom of gauge bosons, we include the ghosts in the loop. The resulting expressions of the matrix elements contain a large number of scalar Feynman integrals, which are reduced to a much smaller set of master integrals (MIs), employing integration-by-parts (IBP) identities~\cite{Tkachov:1981wb,Chetyrkin:1981qh} with the help of \litered~\cite{Lee:2008tj,Lee:2012cn}. The integrals belong to the category of four-point families with two off shell legs of the same virtualities, and these are computed in Refs.~\cite{Gehrmann:2013cxs,Gehrmann:2014bfa} as Laurent series expansion in dimensional regulator $\epsilon$. Employing the results of the MIs, we obtain the FFs~\eqref{eq:FF-defn} to two loops.

The on shell amplitudes in ${\cal N}=4$ sYM are ultraviolet (UV) finite in four dimensions due to vanishing ${\beta}$ function. However, the FFs can exhibit UV divergences if the underlying operator is not SUSY protected which, in the present context, gets reflected by the presence of UV poles in the FFs, ${\cal F}_{\cal K}$, arising from the unprotected Konishi operator ${\cal O}^{\cal K}$. Being a property inherent to the operator, ${\cal O}^{\cal K}$ needs to be renormalized through multiplication of an operator renormalization constant, $Z_{\cal K}$, which reads $\left[{\cal O}_{\cal K}\right]_R=Z_{\cal K} {\cal O}_{\cal K}$.
%
%
$[{\cal O}_{\cal K}]_R$ represents the corresponding renormalized operator. The $Z_{\cal K}$ can be determined~\cite{Anselmi:1996mq, Eden:2000mv, Bianchi:2000hn,Kotikov:2004er, Eden:2004ua,Ahmed:2016vgl} by solving the underlying renormalization group equation and analyzing its Sudakov FFs. The result in terms of its anomalous dimensions, $\gamma_{\cal K}$, is given by $Z_{\mathcal K}  = \exp \left(\sum_{n=1}^{\infty}a^n{2\gamma_{{\cal K},n}}/{n\epsilon}\right)$
%
%
with $\gamma_{{\cal K},1}=-6$ and $\gamma_{{\cal K},2}=24$. For the half-BPS operator, all the anomalous dimensions are identically zero, as guaranteed by the SUSY protection. The UV renormalized FFs are obtained through $\left[{\cal F}_{\lambda}\right]_R=Z^2_{\lambda} {\cal F}_{\lambda}$.
%

The UV finite FFs contain soft and collinear (IR) divergences resulting from the low momentum and/or vanishing angle configurations of the loop momenta. The IR divergences are universal~\cite{Catani:1998bh,Sterman:2002qn,Becher:2009cu,Gardi:2009qi} for an SU(N) gauge theory, which can be expressed as an exponentiation of a quantity containing universal lightlike cusp and collinear anomalous dimensions. The UV renormalized FFs are found to exhibit the expected universal structures of the IR divergences which serve as the most stringent check of our calculation. We find that there is no additional UV divergence from the contact term of two operators, unlike the di-Higgs production through gluon fusion in heavy quark effective theory in QCD~\cite{Zoller:2016iam,Banerjee:2018lfq}. From the perspective of operator product expansion (OPE), in principle, we could expect to encounter additional divergences in the form factors of the two composite operators insertion. The contact terms between two operators are the source of these divergences. Through explicit computation of the FFs, we discover that there is no such contact divergence in the case of double half-BPS or Konishi operators and therefore, we do not need any additional UV counterterm. The absence of contact divergences for the two operator insertion is also found for the production of di-pseudo-scalar in heavy quark effective theory~\cite{Zoller:2013ixa,Bhattacharya:2019oun} and di-Higgs boson in bottom quark annihilation~\cite{H:2018hqz}.


The BDS/ABDK ansatz~\cite{Anastasiou:2003kj,Bern:2005iz}, which says the maximally helicity violating (MHV) amplitude in planar ${\cal N}=4$ sYM is exponentiated in terms of the one-loop result along with the universal anomalous dimensions, gets violated for two-loop six-point amplitudes~\cite{Bern:2008ap,Drummond:2008aq}. In order to capture the deviation from the ansatz, a quantity called finite remainder is introduced~\cite{Bern:2008ap,Drummond:2008aq}.
For the Sudakov FFs of half-BPS operator at two loop~\cite{vanNeerven:1985ja}, both the IR divergence and finite part are found to be exponentiated; however, the finite part stops exhibiting this nature at three loop~\cite{Gehrmann:2011xn}. In order to capture the deviation, following the line of thought for the MHV amplitudes, a finite remainder function (FR) for the FFs at two loop is introduced in Ref.~\cite{Brandhuber:2010ad}, which reads
\begin{align*}
    \mathcal{R}^{(2)}_{\lambda} = 
\mathcal{F}^{(2)}_{\lambda}(\epsilon)-\frac{1}{2}\left(\mathcal{F}^{(1)}_{\lambda}(\epsilon)\right)^2-f^{(2)}(\epsilon)\mathcal{F}^{(1)}_{\lambda}(2\epsilon)-C^{(2)},
\end{align*}
with $f^{(2)}(\epsilon)=-2\zeta_2+\epsilon\zeta_3-\epsilon^2\zeta_2^2/{5}$ and $C^{(2)}={8}\zeta_2^2/{5}$. The quantities $f^{(2)}(\epsilon)$ and $C^{(2)}$ are independent of the number of operators and external states. Representing a two-loop FF in terms of a quantity dictated by the BDS/ABDK ansatz plus an extra part provides a nice way of representing the deviation from the exponentiation --- the ansatz part captures the universal IR divergences that exponentiates whereas the extra part encapsulates the finite part in four dimensions. We compute the FR for both the FFs at two loops and conclude that the finite parts of the two-point FFs with the two operator insertion do not exponentiate, unlike the case of the single half-BPS operator insertion at two loop~\cite{vanNeerven:1985ja}. The results of the form factors and finite remainders are provided as \ancillary~ files with
the \arXiv~(version-1) submission. 

\begin{figure}
\includegraphics[width=6.8cm, height=10cm]{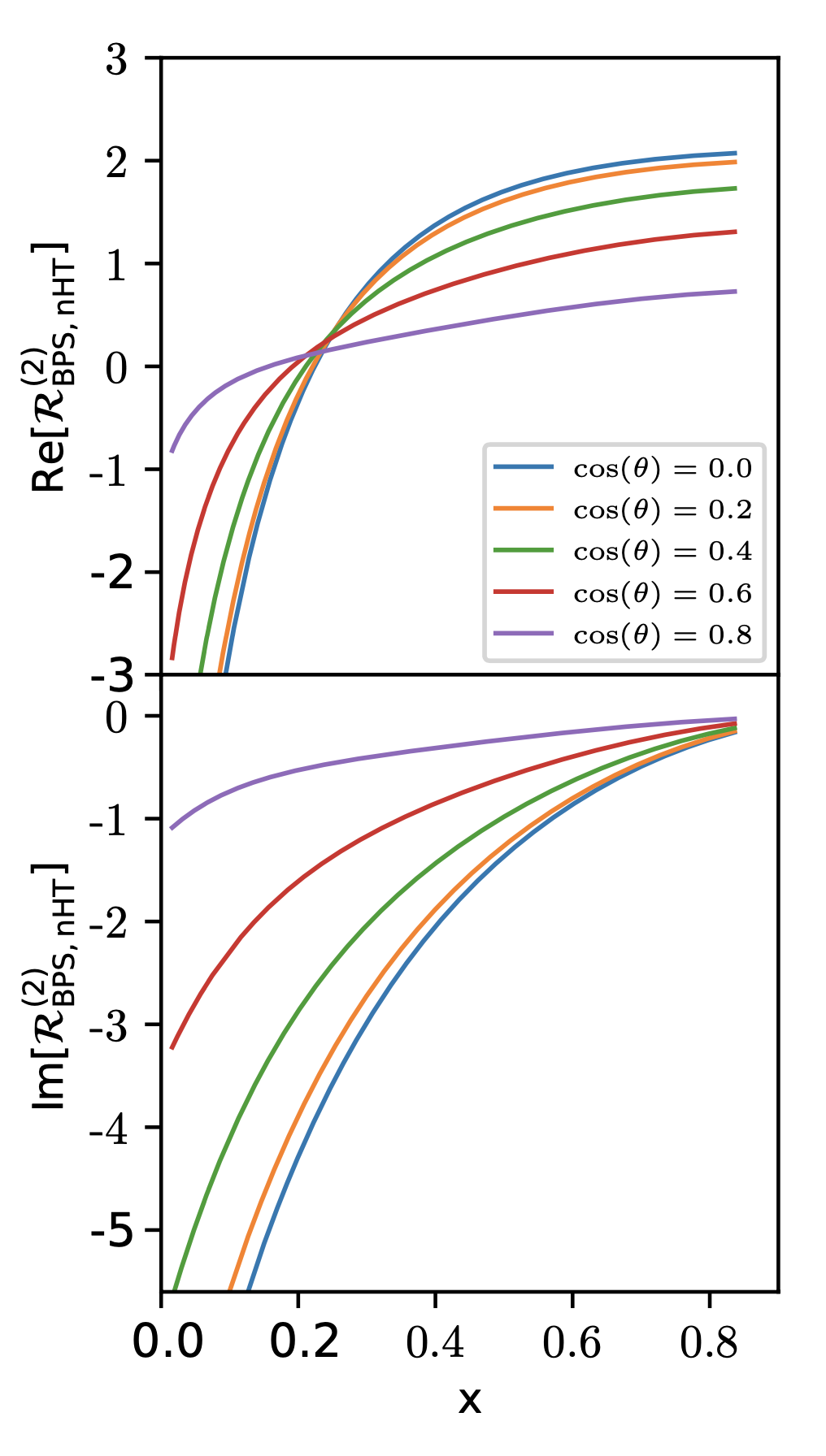}
\caption{\label{fig:epsart} Next-to-highest transcendental terms in the two-loop finite remainder of the double half-BPS operator}
\label{fig:nHTBPS}
\end{figure}


\section{III.\,\,PRINCIPLE OF UNIFORM TRANSCENDENTALITY}
\label{sec:UT}

It is a general belief, albeit based on observations, that the FFs and FRs of a SUSY protected operator, such as half-BPS, exhibit the behavior of UT; i.e., they contain only HT weight terms --- commonly known as the principle of uniform transcendentality (PUT). No deviation from this conjecture has ever been found. In this article, for the first time, we report that the property of UT does not extrapolate to the FFs of double insertions of a SUSY protected operator. We find that though the FF of half-BPS primary is UT at one loop, it no longer holds true at two-loop,
\begin{align}
\label{eq:LT-BPS}
    &{\cal F}_{{\rm BPS},{\rm nHT}}^{(1)}=0\,, \qquad\quad 
    {\cal F}_{{\rm BPS},{\rm nHT}}^{(2)} \neq 0\,,
\end{align}
where ${\rm nHT}$ represents the next-to-highest-transcendental terms. Remainder functions also obey same property. Therefore, the property of UT for SUSY protected operator can not be generalized to a more general class of FFs with more than the one operator insertion. To be more specific, among the nHT terms at two loop, only the transcendental 3 term is nonzero; the remaining lower ones identically vanish,
\begin{align}
\label{eq:LT-BPS}
    {\cal F}_{{\rm BPS},{\rm nHT}}^{(2)} = {\cal F}_{{\rm BPS}}^{(2),\tau(3)}\,, \qquad  {\cal F}_{{\rm BPS}}^{(2),\tau(<3)}=0\,,
\end{align}
%
where the FFs are written as ${\cal F}_{\lambda}^{(n)}=\sum_{l=0}^{2n}{\cal F}_{\lambda}^{(n),\tau(l)}$. The $\tau(l)$ represents the terms with transcendentality weight $l$. Since the result is too big to be presented here, we provide a graphical presentation of the nHT terms of the FR, ${\cal R}_{{\rm BPS},{\rm nHT}}^{(2)}$, in Fig.~\ref{fig:nHTBPS} 
to demonstrate the dependence on scaling variables $x$ and y. We plot the real (Re) and imaginary (Im) parts as a function of the Landau variable $x$ for different choices of  $\rm cos\theta$, where $\theta$ is the angle between one of the particles corresponding to the half-BPS operator and one of the initial state scalars in their center of mass frame, defined through $2 m_\lambda^2y = (s-2m_\lambda^2-\cos\theta\sqrt{s(s-4m_\lambda^2)})$.
The FR is also seen to be invariant under $\rm cos\theta$ $\leftrightarrow$ $-\rm cos\theta$, as expected for a purely bosonic scattering. Since this symmetry is not used in the setup of the calculation, this serves as a strong check on the finite part of the results. For numerical evaluation of the polylogarithms, we make use of the  package from Ref.~\cite{Frellesvig:2016ske}.


In contrast to the form factor of half-BPS, the UT is not a property for the two-point FFs with the single insertion of unprotected Konishi operator, which is verified to three loops~\cite{Nandan:2014oga,Ahmed:2016vgl}. The FFs with double insertions exhibit the behavior consistent with this expectation,
\begin{align}
\label{eq:LT-BPS}
    &{\cal F}_{{\cal K}, \,{\rm nHT}}^{(1)}\neq 0\,, \qquad\quad 
    {\cal F}_{{\cal K}, \,{\rm nHT}}^{(2)} \neq 0\,.
\end{align}

\begin{figure}
\includegraphics[width=6.8cm, height=10cm]{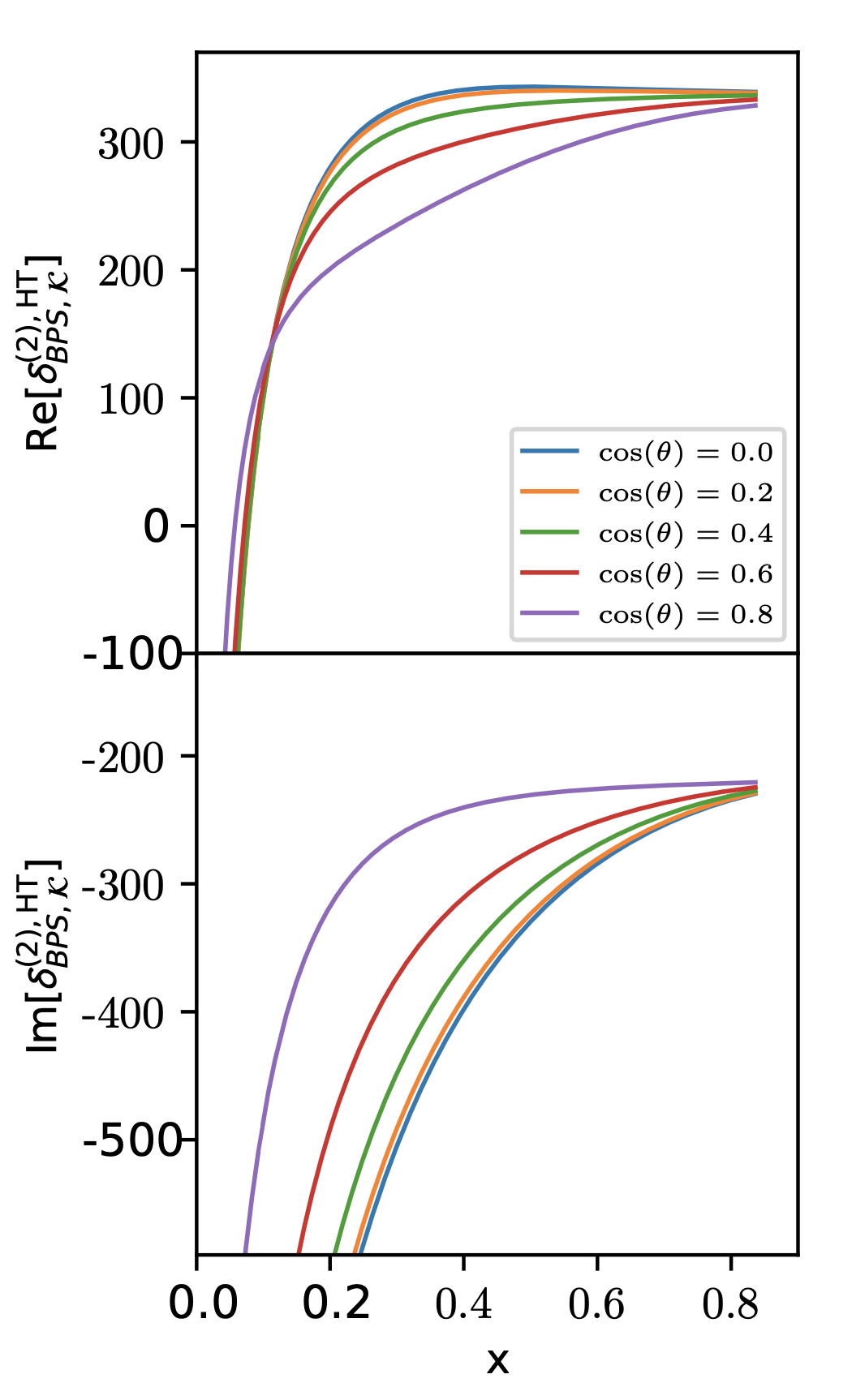}
\caption{\label{fig:epsart} Difference ($\delta$) between the highest transcendental terms of two-loop finite remainders of double half-BPS and Konishi}
\label{fig:dBPSK}
\end{figure}

In Ref.~\cite{Loebbert:2015ova}, in the context of FFs with the one operator insertion, it is conjectured that the HT weight parts of every two-point minimal FF, including that of Konishi, are identical to that of half-BPS. Through our computation, for the first time, we report the deviation from this conjecture, in particular, this property fails to hold true for the double insertions of operators. Our findings show that
\begin{align}
\label{eq:LT-BPS}
    &{\cal F}^{(1)}_{{\cal K}, \,{\rm HT}} \neq {\cal F}_{{\rm BPS}, \,{\rm HT}}^{(1)}\,, \qquad\quad
    {\cal F}^{(2)}_{{\cal K}, \,{\rm HT}} \neq {\cal F}_{{\rm BPS}, \,{\rm HT}}^{(2)}\,.
\end{align}
Hence, the conjecture fails, in general, to be extrapolated for the case involving the two operator insertion. 
In addition to the analytical check, we verify it numerically which is displayed graphically through  Fig.~\ref{fig:dBPSK}.


\section{IV.\,\,Principle of Maximal Transcendentality}
\label{sec:PMT}

The conjecture of PMT establishes a bridge between ${\cal N}=4$ sYM and QCD. It states that the HT terms of certain quantities in ${\cal N}=4$ sYM and QCD are identical upon converting the fermions in QCD from fundamental to adjoint representation through $C_A=C_F=2 n_f T_F=N$, where $C_A$ and $C_F$ are the Casimirs in adjoint and fundamental representations, respectively, $n_f$ is the number of light quark flavors, and $T_F$ is the normalization factor in fundamental representation. 
%
%
%
%
The conjecture is found to hold true to three loops for two-point FFs with the one operator insertion while comparing the quark/gluon FFs in QCD and that of a half-BPS primary. The question arises if the PMT carries over to a more general class of FFs and correlation functions. Through our computations, for the first time, we find that the PMT does not hold true for the FFs with the two operator insertion. The HT terms of two-point FFs of the double half-BPS operator do not match with that of the di-Higgs produced either through gluon fusion~\cite{Banerjee:2018lfq} or through bottom quark annihilation~\cite{H:2018hqz},
\begin{align}
\label{eq:diff-diHiggs}
	\mathcal{F}_{\rm BPS,HT}^{(n)} \neq \mathcal{F}_{gg\rightarrow HH,{\rm HT}}^{(n)} \neq \mathcal{F}_{b\bar b\rightarrow HH,{\rm HT}}^{(n)}\,.
\end{align}
%
Consequently, the conjecture of PMT can not be extrapolated to the general class of FFs and correlation functions. 

\section{V.\,\,Regge and Collinear Limit}
\label{sec:RClimit}

Though the PUT and PMT do not hold true for the double half-BPS operator, we intend to see whether it can be restored in some kinematic limit. Keeping this in mind, we investigate the behavior of the nHT terms analytically with the help of \polylogtools~\cite{Duhr:2019tlz} in the kinematic regime captured by x$\rightarrow$ 0 and $\cos{\theta} \rightarrow$ 1. The former one corresponds to the Regge limit whereas the latter makes one of the color and colorless particles collinear, and thereby effectively converting it into a three point scattering. Upon taking the limits simultaneously, the entire nHT term is found to vanish for the double-BPS form factor at two loop, which implies the restoration of the PUT,
\begin{align}
   \lim_{\substack{x \rightarrow 0 \\ \cos\theta \rightarrow 1}} {\cal F}_{{\rm BPS},{\rm nHT}}^{(2)}  = 0\,.
\end{align}
Moreover, the PMT also gets restored to two loops only for the di-Higgs boson production through gluon fusion,
\begin{align}
\label{eq:diff-diHiggs-RClimit}
\lim_{\substack{x \rightarrow 0 \\ \cos\theta \rightarrow 1}}	\Big[ \mathcal{F}_{gg\rightarrow HH,{\rm HT}}^{(n)} =  \mathcal{F}_{\rm BPS,HT}^{(n)} \neq \mathcal{F}_{b\bar b\rightarrow HH,{\rm HT}}^{(n)} \Big] \,.
\end{align}
We also find that in this limit the difference between the HT terms of half-BPS and Konishi FFs disappears, and thereby the conjecture of having identical HT weight parts of every two-point minimal FF is also reinstated.

The underlying reason behind the transcendentality principles is not well-understood till date. From the perspective of OPE, the generic expansion of the correlator of the double operator insertion should contain not only SUSY protected but also unprotected operators, and therefore, {\it a priori} there is no reason to believe the existence of transcendentality principles for FFs involving multiple operator insertion. Either the OPE or an explicit computation of the FFs can only reveal the true nature. Our experience also shows that the transcendentality principles for the minimal FFs can be correlated to the existence of a complete factorization of the leading order amplitude at any loop order under consideration. For the form factor of a double operator insertion, the leading order amplitude does not factorize from the complete result either at one or two loop. Consequently, the transcendentality principles, in general, fail to hold true for the double operator insertion, which however gets reinstated in the simultaneous Regge and collinear limit. It leads us to the understanding that essentially the process dependence is embedded into the leading order result, and therefore, upon a successful factorization of the leading order amplitude at certain loop order, we end up with a quantity exhibiting universal nature from the perspective of transcendentality. 
\section{VI.\,\,Conclusions}
For the first time, we present the form factors with the insertions of two identical local gauge invariant operators to two loops in ${\cal N}=4$ sYM theory by performing a state-of-the-art computation. In particular, we compute the scalar FFs with double insertions of half-BPS primary and Konishi operators. 
Through this calculation, we take a step forward to go beyond the FFs of single operator insertion and enter into the domain of a more general class of FFs. To validate our computations, we check the infrared poles, which agree with the predictions. Moreover, the appearance of expected kinematic symmetry inherent to the bosonic FFs provides a strong check on the finite parts of our results. 

The findings enable us to reach a number of important conclusions. For the first time, the conjecture that the FFs of SUSY protected operators are always UT is found not to hold true at two loop for the FFs of the double operator insertion, in sharp contrast to the Sudakov FFs. In particular, though the FF of the double half-BPS primary is UT at one loop, it fails to exhibit this property at two loop. In accordance with our expectation, we find the FFs of a SUSY unprotected operator, Konishi, to be not UT. From our experience, we find that the factorization of the leading order amplitude at any loop order can be accounted for the existence of UT property.

The conjecture that the highest transcendentality weight terms of every two-point minimal FF are identical to that of half-BPS is also found not to hold true for the two operator insertion. In other words, the HT weight terms of unprotected Konishi are not identical to that of half-BPS both at one and two loop. 

We find that the principle of maximal transcendentality, which says the highest weight terms of quark/gluon FFs in QCD are identical to that of scalar FFs of half-BPS primary, can not be extrapolated to the case of the two operator insertion. The HT weight terms of the double half-BPS FFs do not match with that of di-Higgs production through gluon fusion or bottom quark annihilation in QCD, observed for the first time in this context. 


By computing the finite remainder function at two loop, we confirm that the finite parts of the FFs of double half-BPS do not exponentiate, in contrast to the corresponding FFs of single half-BPS at two loop~\cite{vanNeerven:1985ja}. 
Moreover, we discover that there is no contact divergence in the case of double half-BPS or Konishi operators, and therefore, we do not need any additional UV counterterm. Our work, in addition to providing a better understanding of the nature of generalized FFs, opens the door for further analytic calculations of a general class of FFs.


\section*{ACKNOWLEDGEMENTS}
We are grateful to S. Seth for providing his python code to incorporate Majorana fermions in \qgraf. We sincerely thank G. Yang, A. von Manteuffel, and A. Tumanov for their comments on the manuscript. We also thank G. Yang for discussions on the operator product expansion. We wish to acknowledge discussions about Regge and collinear limits with A. von Manteuffel. PB thanks A. Signer for discussions. TA thanks S. Zoia for linguistic advice. The work of TA received funding from the European Research Council (ERC) under the European Unions
Horizon 2020 research and innovation programme, \textit{Novel structures in scattering amplitudes} (Grant Agreement No. 725110). 

\bibliography{main} 
\end{document}